\documentclass[preprint]{elsarticle}

\usepackage{lineno,hyperref}
\modulolinenumbers[5]

\journal{Speech Communication}

\usepackage{graphicx}
\usepackage{multirow}
\usepackage{pdfpages}
\usepackage{amsmath}

\biboptions{authoryear}

\begin{document}

\begin{frontmatter}

\title{Improving speaker de-identification with functional data analysis of f0 trajectories} 

\author[1,2]{Lauri Tavi\fnref{myfootnote}}
\author[2]{Tomi Kinnunen}
\author[2,3]{Rosa González Hautamäki}
\fntext[myfootnote]{Corresponding author.
Address: Yliopistokatu 4, Agora, Joensuu, Finland.
Email: lauri.tavi@uef.fi}
\fntext[myfootnote1]{DOI: \url{https://doi.org/10.1016/j.specom.2022.03.010}, \emph{Accepted date}: 11 March 2022. This manuscript version is made available under the \href{https://creativecommons.org/licenses/by-nc-nd/4.0/}{CC-BY-NC-ND 4.0 license}
\vspace{-1cm}}

\address[1]{School of Humanities, University of Eastern Finland, Finland}
\address[2]{School of Computing, University of Eastern Finland, Finland}
\address[3]{Electrical and Computer Engineering, National University of Singapore, Singapore}

\begin{abstract}
Due to a constantly increasing amount of speech data that is stored in different types of databases, voice privacy has become a major concern. To respond to such concern, speech researchers have developed various methods for speaker de-identification. The state-of-the-art solutions utilize deep learning solutions 
which can be effective but might be unavailable or impractical to apply for, for example, under-resourced languages. Formant modification is a simpler, yet effective method for speaker de-identification which requires no training data. Still, remaining intonational patterns in formant-anonymized speech may contain speaker-dependent cues. This study introduces a novel speaker de-identification method, which, in addition to simple formant shifts, manipulates f0 trajectories based on functional data analysis. The proposed speaker de-identification method  will conceal plausibly identifying pitch characteristics in a phonetically controllable manner and improve formant-based speaker de-identification up to 25\%.
\end{abstract}


\end{frontmatter}


\section{Introduction}

\noindent Speech is a biometric identifier and therefore considered as personal information \citep{nautsch2019gdpr}. People can be identified from a constantly increasing amount of speech databases, which has raised concerns about voice privacy during recent years. Human listeners can recognize speakers’ identities reliably only within a small speaker population \citep{farrus2018voice}, but automatic speaker recognition systems, which are not restricted to small speaker populations, typically outperform naive listeners in unfamiliar speaker recognition tasks \citep{hautamaki2010approaching}.

The recognition accuracy of both human and even state-of-the-art automatic systems is known to decline when speakers deliberately change their speech characteristics to conceal their identities, i.e. \emph{disguise} their voices \citep{hautamaki2017acoustical}. It can be assumed, however, that few people have disguised their voices in speech databases, which are collected by different operators, such as officials, scientists, reporters, commercial service providers --- and potentially even criminals. To regulate collection practices, in 2016 the European Union passed a well-known regulation, the General Data Protection Regulation (GDPR), to defend natural persons’ right to the protection of their personal data \citep{GDRP}. Still, different operators may have varied practises in the usage of speech data, which can be either approved or not by the persons whose voice data is being processed. Ethical concerns can also arise in situations, where speech data is sent to a third party without the consent of speech donors. For instance, in phonetic studies, performing forced alignment or automatic speech recognition (ASR) might require uploading speech recordings to web services, which might compromise speakers’ right to privacy.

As a remedy to the voice privacy issues, speech researchers have begun developing \emph{speaker de-identification} techniques. Ideally, these techniques should preserve the intelligibility and naturalness of speech but suppress speaker-related acoustic characteristics so that the anonymized speech cannot be linked back to the original speaker \citep{tomashenko2020introducing}. Speaker de-identification, or speaker anonymization, is achieved by modifying selected speech properties so that speakers become unrecognizable either by automatic systems, listeners, or both. One of the simplest speaker de-identification methods is to manipulate speakers’ pitch using, for instance, \emph{pitch-synchronous overlap-and-add} (PSOLA) \citep{moulines1990pitch}, although this is not sufficient to protect privacy against state-of-the-art automatic speaker verification (ASV) systems \citep{patino2020speaker}. It is still well-known that speakers' individual pitch characteristics contain speaker-related information \citep{rose2002forensic,adami2007modeling} since people have physiological differences in their larynges and individual tendencies to adopt different intonational styles.

The current mainstream approach for speaker de-identification is to perform ASR, followed by speech synthesis or voice conversion \citep{magarinos2017reversible,bahmaninezhad2018convolutional}. The state-of-the-art techniques use speech synthesizers based on \emph{x-vector speaker embeddings} \citep{snyder2017deep} and \emph{bottleneck features}. X-vector embeddings are assumed to represent speaker-related information while bottleneck features gather linguistic content-related information \citep{srivastava2020design}. The f0 values are also extracted from the source speech, which are then directly used or modified for anonymized synthesis \citep{champion2021study}. As a result, resynthesizing speech content with other than original speaker x-vector embeddings anonymizes speech efficiently. As a downside, such systems contain a large number of model parameters and require either training or adaptation data. For under-resourced languages, such data (or even a pre-trained model) might be unavailable.

A simpler speaker de-identification method which requires no training data is to shift formant frequencies using, for example, a predefined coefficient. This results in a phonetically transparent, data-free and lightweight approach, at the expense of less effective speaker de-identification performance  \citep{patino2020speaker}. Besides constant shifts, formants can be also modified in a data-driven manner by using formant statistics of desired speakers, as done by \cite{dubaguntaadjustable}. Because formants correspond to resonances of the vocal tract, and lower formant frequencies are associated with longer vocal tract length \citep{lammert2015short}, decreasing formant frequencies will cause a speaker to sound as if she/he has a longer vocal tract and deeper voice. Contrariwise, increasing formant frequencies will give an impression of a shorter vocal tract and a brighter voice. The scope of physically plausible formant shifts for natural sounding speech is, however, somewhat limited. One might be able to restore the original formants after a reasonable number of attempts. Therefore, adding other speech modifications might be necessary to protect voice privacy.

The aim of this study is to investigate a phonetically controllable approach to speaker de-identification, which modifies speakers’ vocal tract and voice source features, with a special focus on \emph{f0 trajectories}. To modify vocal tract features, a simple shift of the first three formant frequencies is applied. For manipulation of f0 trajectories, a more sophisticated approach leveraging \emph{functional data analysis} or FDA \citep{ramsay2009introduction} is proposed. In FDA, discrete (sampled) data is first transformed into a continuous domain through appropriate functionals (such as splines), and then analysed using specific statistical methods developed to handle such data 
The particular method of interest is \emph{functional principal component analysis} (fPCA), a continuous-domain counterpart of conventional PCA. The main idea behind FDA methods is to capture dynamic changes in time series data instead of traditional statics. Because in speech analysis dynamic changes are often more relevant than global statistics, FDA techniques have been successfully applied to phonetic data during recent decades \citep{gubian2009functional,gubian2010automatic,zellers2010redescribing,gubian2011joint,gubian2015using,cronenberg2020dynamic}.

As far as the authors are aware, FDA methods have not been previously applied to speaker de-identification. In this study, f0 trajectories of anonymized speech are reconstructed based on functional PC weightings of 1) disguised and 2) cross-sex speech. Because disguised speech is known to substantially deteriorate ASV system performance \citep{gonzalez2016age,hautamaki2017acoustical}, this study aims to discover the main pitch variations related to disguised voices and utilize this knowledge in anonymization of speakers' f0 trajectories. Similarly, to demonstrate a more practical approach for applying FDA to speaker de-identification, anonymization of f0 trajectories was also implemented utilizing natural pitch differences across sexes.

In general, speaker de-identification must fulfil at least two requirements: 1) it provides protection against speaker recognition and 2) preserves speech intelligibility. In this study, the privacy protection of the proposed speaker de-identification techniques is tested using an x-vector-based ASV system. The intelligibility of anonymized speech is evaluated using an objective intelligibility measure. Next, Section \ref{mame} will introduce the speech corpus and f0 analyses. Section \ref{SDM} will describe the de-identification techniques in detail, while validation procedures for anonymized speech are introduced in Section \ref{vali}. Finally, the results concerning speaker de-identification and speech intelligibility are presented in Section \ref{Res} and discussed in Section \ref{Diss}.

\section{Speech materials and fPCA modelling}
\label{mame}

\subsection{AVOID corpus} \label{speechdata}

\noindent This study aims to exploit pitch characteristics from actual disguised and cross-sex speech to improve speaker de-identification. To fulfil this aim, the Age-related VOIce Disguise (AVOID) corpus \citep{gonzalez2016age} is adopted. Unlike previous corpora used in other speaker de-identification studies --- such as the data sets provided by the recent VoicePrivacy challenge 2020 \citep{tomashenko2020introducing} --- AVOID contains naturally disguised voices besides modal speech. This allows modelling of disguised voices and comparison of proposed speaker de-identification methods against \emph{human-based} de-identification, for reference purposes. 

The AVOID corpus was recorded in 2015 with three recording devices at a sampling rate of 44.1 kHz. It includes a total of 60 Finnish speakers (31 female and 29 male), which produced 13 different sentences (Finnish versions of ``The rainbow passage'' and ``The north wind and the sun'', along with two selected TIMIT sentences in English). The sentences were spoken in three different styles: \emph{modal}, \emph{intended elderly} and \emph{intended child}, recorded twice in two different sessions. In this study, only microphone-recorded (Glottal Enterprises M80) speech is used, resulting in a total of 3600 (60 speakers $\times$ 10 utterances $\times$ 3 speaking styles $\times$ 2 sessions) utterances in Finnish. In order to keep the focus on voice disguise and cross-sex variation, we removed plausible effects of accented speech and overly short utterances by excluding the two TIMIT utterances and one short Finnish utterance (numbered as ``13'', ``12'', and ``05'', respectively). 

\subsection{f0 extraction and smoothing} \label{f0extract}

\noindent f0 trajectories were extracted using the autocorrelation method available in Praat \citep{Praat} and PraatR \citep{albin2014praatr}. 
Because the speakers in the AVOID corpus have unusually wide pitch ranges mainly due to intended child speech, it was necessary to test different pitch settings to find optimal floor and ceiling values. After testing different values on the whole corpus, pitch floor and ceiling were set, respectively, to 140 Hz and 520 Hz for female speakers. The corresponding settings for male speakers were 65 Hz and 380 Hz.

After extracting f0 trajectories, they were interpolated to fill unvoiced segments and converted into semitone units. Each f0 trajectory was then transformed to a continuous-domain curve with \emph{B-spline} functions. For the spline basis system, the order was set to four and the number of basis functions to 202, which corresponds approximately to one third of the normalized number of f0 samples. It is a standard practise to use less interior breakpoints, or "knots", than original data samples to avoid overfitting or even replicating measurement errors. Lambda parameter, which specifies the amount of smoothing, was set to 1e-8.
Like with the selection of the pitch parameters, we tested different smoothing parameters on the whole data and visually inspected the resulting f0 curves in order to model appropriately smoothed curves. We followed the guidelines presented by Gubian, Torreira, and Boves (2015); in their study, the effect of choosing different parameters for a B-spline system is well demonstrated. Neither landmark registration nor dynamic time warping were applied to F0 curves, as the aim is to discover global pitch changes in disguised and cross-sex speech in a text-independent manner, rather than analysing local phonetic phenomena behind specific linguistic units. To carry out FDA in R \citep{R123}, this study used \texttt{fda} \citep{fda} and \texttt{fda.usc} \citep{fda_usc} packages.

\begin{figure} [h!]
\centering
\includegraphics[width=0.9\textwidth, angle=0]{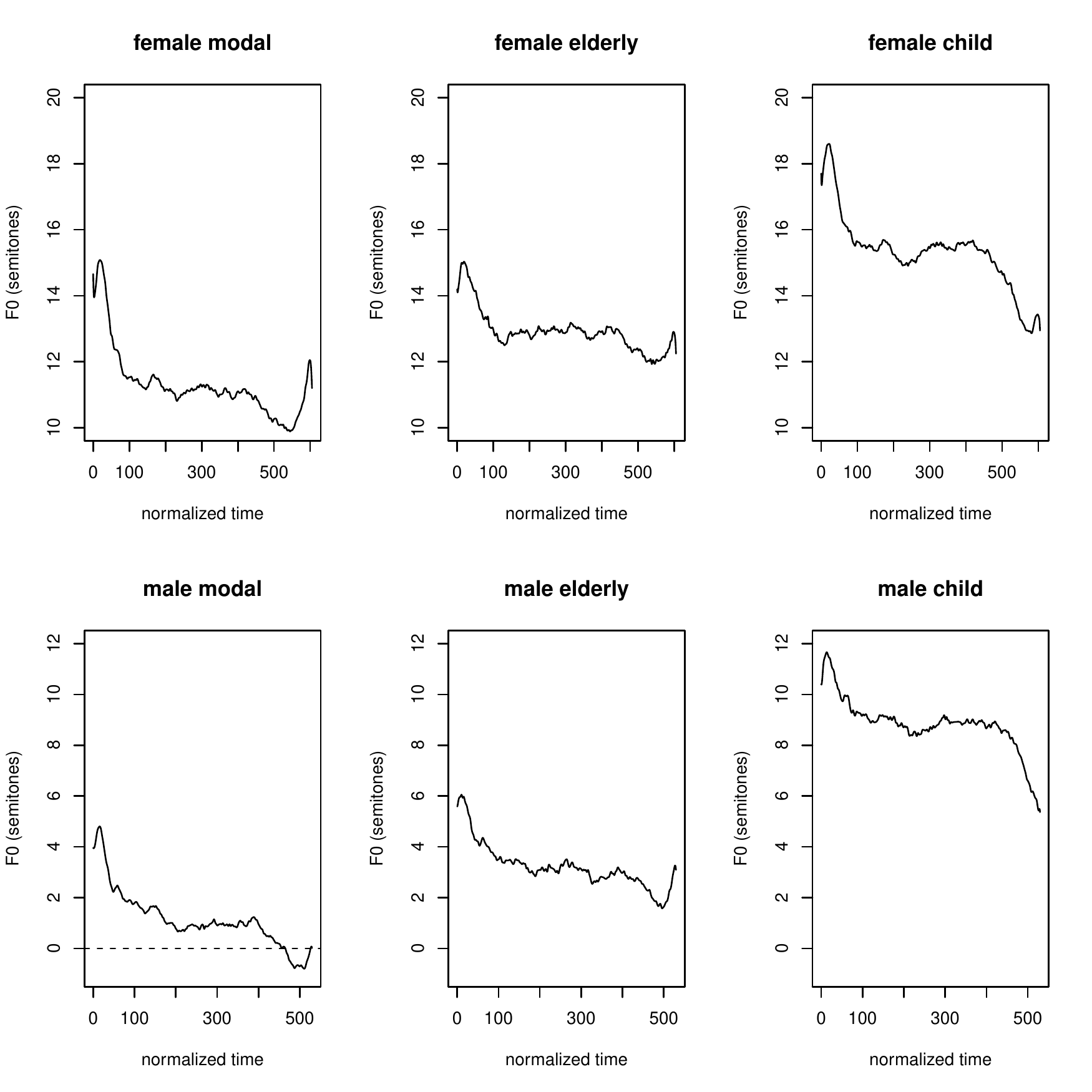}
\caption{Mean f0 curves for female (above) and male (below) speakers. The curves on the left present modal f0 trajectories and the curves on the right present intended child f0 trajectories. The curves in the middle show intended elderly f0 trajectories.}
\label{fig: fig1}
\end{figure}

Figure \ref{fig: fig1} displays functional \emph{mean} f0 curves of all female and male speakers in the three voice conditions. As can be seen, f0 curves of disguised voices are located at higher frequencies compared to modal ones and modal and intended child voices differ most for both sexes. No major differences exist between mean curve shapes: all the three voice conditions have a high peak at the beginning, followed by a general descending trend, although the intended child voices have a flat and long ascent at the second half of the curve. Similarly, the main difference between modal female and male curves is the elevation rather than the overall shape. Yet, compared to female speakers, male speakers have a greater elevation of mean f0 curve when disguising their voices using child-like speech. For both sexes, the mean f0 curves for intended child voices exhibit a distinguishable voice disguise strategy, which can partially explain the negative effect of child-like speech on ASV accuracy \citep{gonzalez2016age}.

\subsection{Functional principal component analysis} \label{fPCA}

\noindent One of the most used methods in FDA is \emph{fPCA} \citep{ramsay05}, a continuous-domain counterpart of conventional PCA. The key difference of PCA and fPCA is that the latter uses basis \emph{functions} rather than basis \emph{vectors} to represent data. The vector space basis used in PCA --- a set of \emph{eigenvectors} --- is replaced by a set of \emph{eigenfunctions}, each associated with a corresponding eigenvalue \citep{ramsay2009introduction}. In speech analysis, fPCA renders a model for input curves, such as f0 or formant trajectories, as a set of mean and principal curves (PCs) and their weights, also known as \emph{scores} \citep{gubian2015using}. The model for input curves is formally presented by
    \begin{equation}\label{eq:f0-finite-expansion}
        f(t) \approx \mu(t) + \sum_{i=1}^n s_i \cdot \text{PC}_i(t),
    \end{equation}
where an approximation of a continuous curve $f(t)$ with domain $0 \leq t \leq 1$ is reconstructed as the sum of a mean curve $\mu(t)$, a set of $n$ principal component curves $\{\text{PC}_i(t) \}$ and their weights $\{s_i\}$. The more PC curves and their weights are used, the more similar the reconstructed curves will be with the original curves. 

\begin{figure} [h!]
\centering
\includegraphics[width=1.0\textwidth, angle=0]{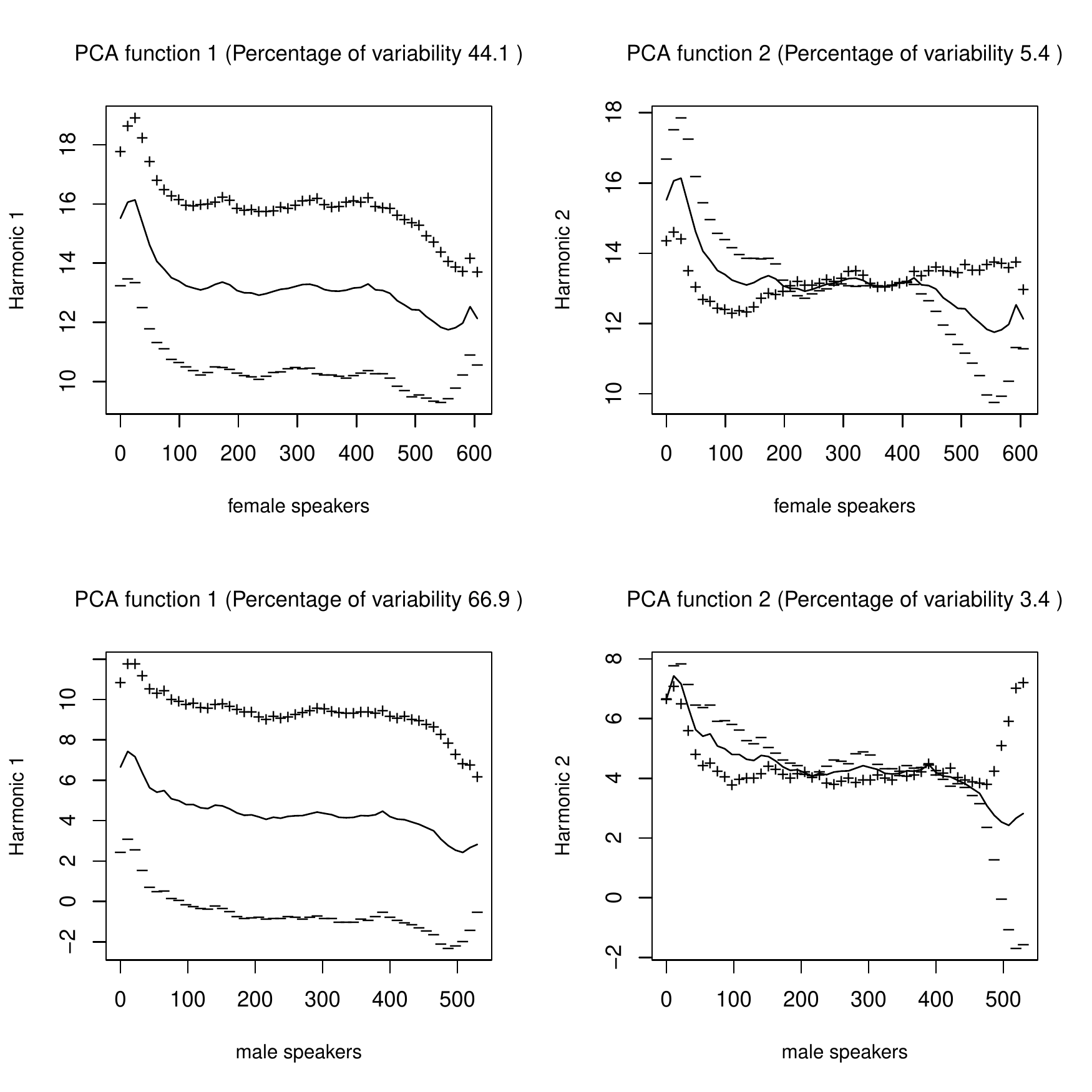}
\caption{Traditional demonstration of the effect of PC1 (left) and PC2 (right) curves on the mean curve (solid curve). Standard deviation of weightings of PCs are multiplied with PCs and either added to or subtracted from the mean curve to form plus (marked as “+ + +”) and minus (marked as “- - -“) curves. PC functions are presented separately for female speakers (above) and for male speakers (below).}
\label{fig: fig2}
\end{figure}
Three separate fPCAs were applied to f0 curves extracted from the AVOID corpus: two fPCAs for finding the primary modes of variation related to voice disguise conditions for female and male speakers, and one fPCA for finding the primary modes of variation within modal cross-sex speech.

Figure \ref{fig: fig2} presents f0 variation in the voice conditions captured by PC1 and PC2. For female speakers (above), PC1 explains 44\% of pitch variation (on left), while PC2 explains only 5\% of variation (on right). For male speakers (below), the corresponding variability percentages are 67 (on left) and 3 (on right). Solid lines show mean curves for all voice conditions, and plus and minus curves are obtained by multiplying PC1 or PC2 with standard deviation of their weightings, which is then either added to or subtracted from the mean curve. The purpose of the plus and the minus curves is to present their effect on the mean curve as curve modifiers.

\begin{figure} [h!]
\centering
\includegraphics[width=0.7\textwidth, angle=0]{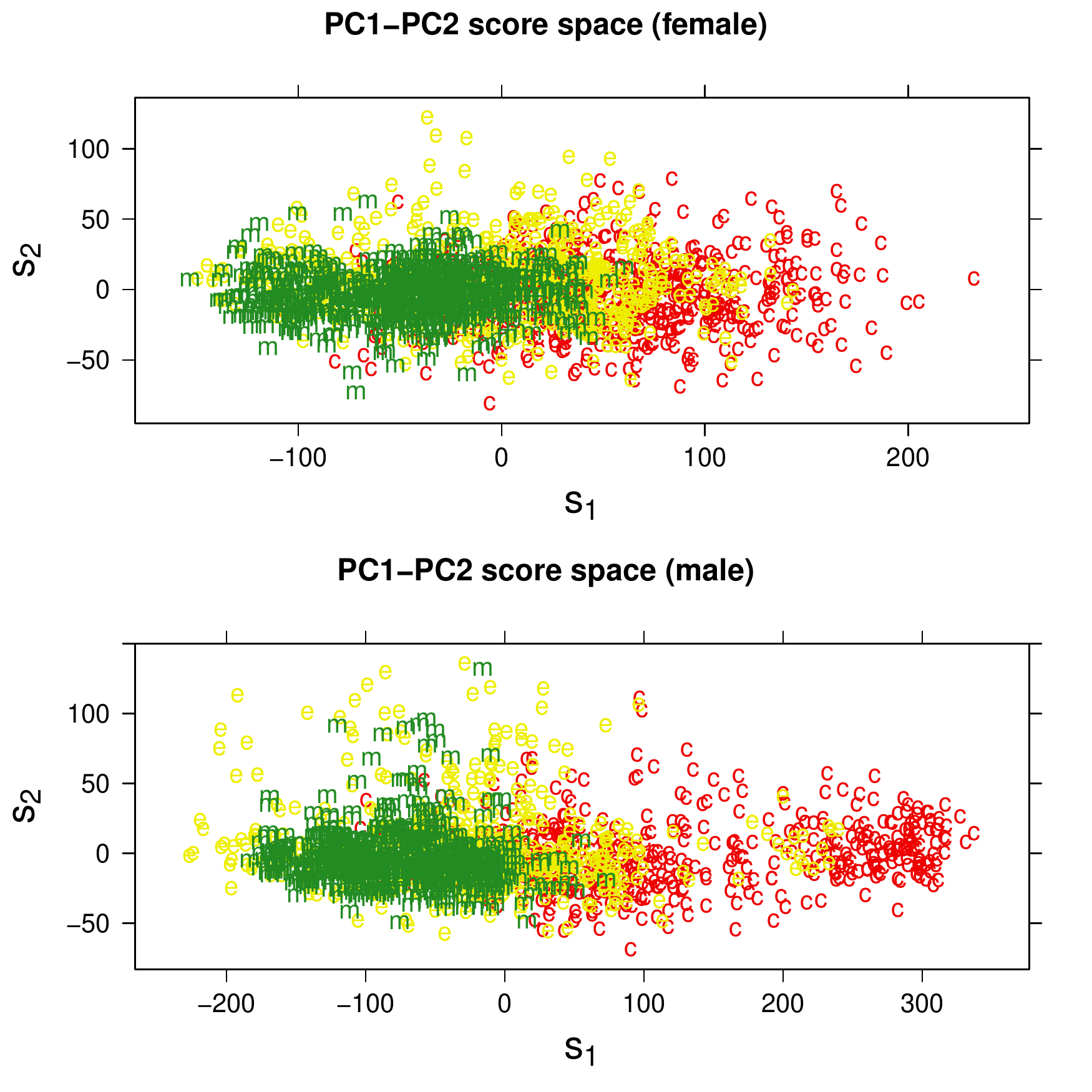}
\caption{Female (above) and male (below) speakers’ PC1--PC2 score spaces. Letters \emph{m}, \emph{e} and \emph{c} indicate PC scores (i.e. the $s_1$ and the $s_2$) calculated from modal, elderly and child voices, respectively.}
\label{fig: fig3}
\end{figure}

Figure \ref{fig: fig2} suggests that for both sexes PC1 primarily alters the position of the f0 curve while PC2 is related to the tilt of the curve. Moreover, comparison of Figure \ref{fig: fig2} and Figure \ref{fig: fig1} indicates that the plus curves constructed using PC1 and its weights are similar compared to the mean f0 curves of child voices in Figure \ref{fig: fig1}. This indicates that PC1 captures well the effect of intended child voice on pitch variation. This observation is supported also by Figure \ref{fig: fig3}, which shows scatter plots of the $s_1$ and the $s_2$ for female (above) and male (below) speakers.

Figure \ref{fig: fig3} illustrates how the three voice conditions are grouped based on the first two principal weightings. Even though the weightings of intended elderly voices (\emph{e} letters) overlap with the other two groups of weightings (\emph{m} and \emph{c} letters), the weightings for modal and intended child voices are visibly separated. The difference between the groups, however, mainly relies on $s_1$: the $s_1$ of modal voice is more associated with negative values while $s_1$ of the intended child voice is associated with mostly positive values. 

\begin{figure} [h!]
\centering
\includegraphics[width=0.8\textwidth, angle=0]{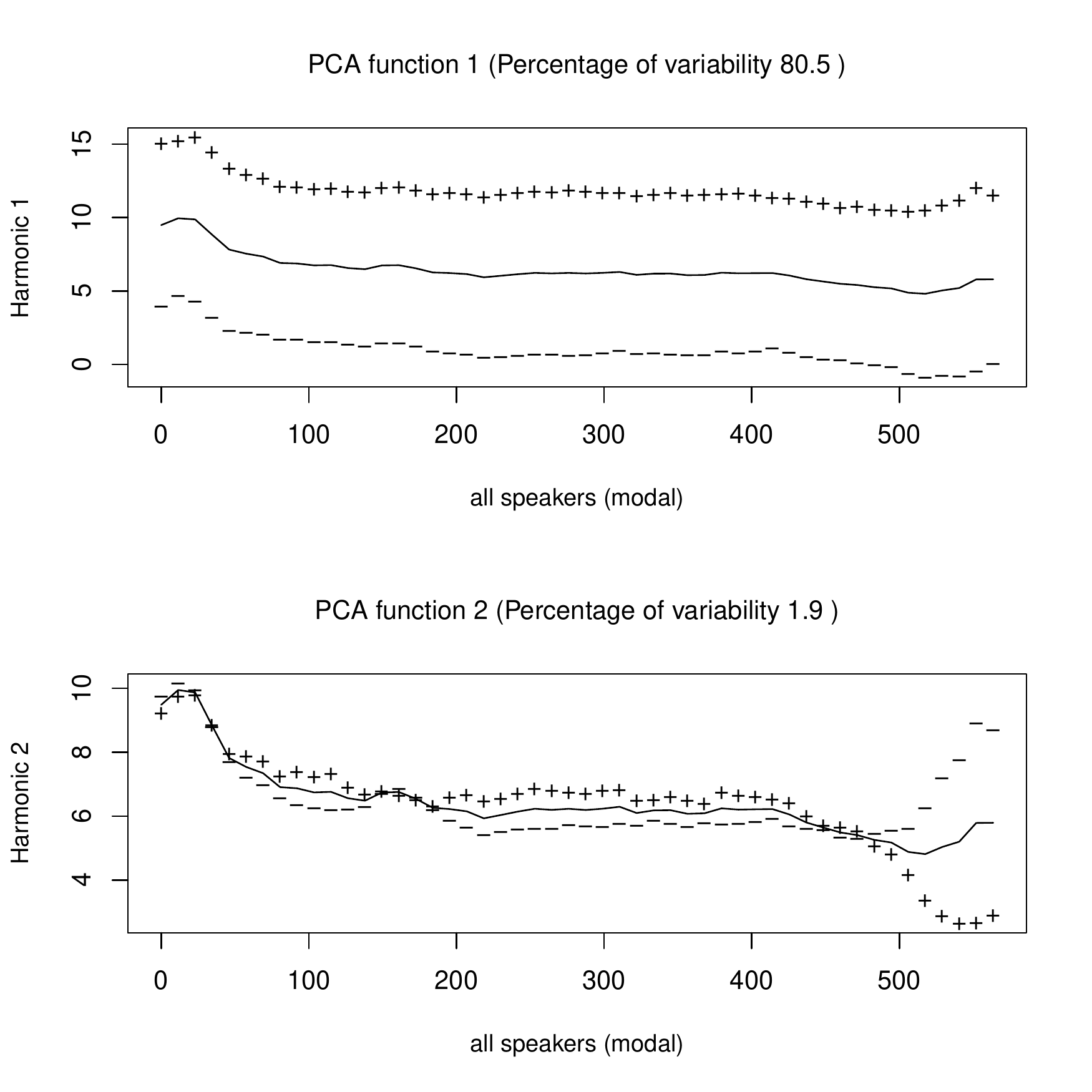}
\caption{Traditional demonstration of the effect of PC1 (above) and PC2 (below) curves on the mean curve (solid curve). Standard deviation of weightings of PCs are multiplied with PCs and either added to or subtracted from the mean curve to form plus (marked as “+ + +”) and minus (marked as “- - -“) curves.}
\label{fig: fig4}
\end{figure}

\begin{figure} [h!]
\centering
\includegraphics[width=0.7\textwidth, angle=0]{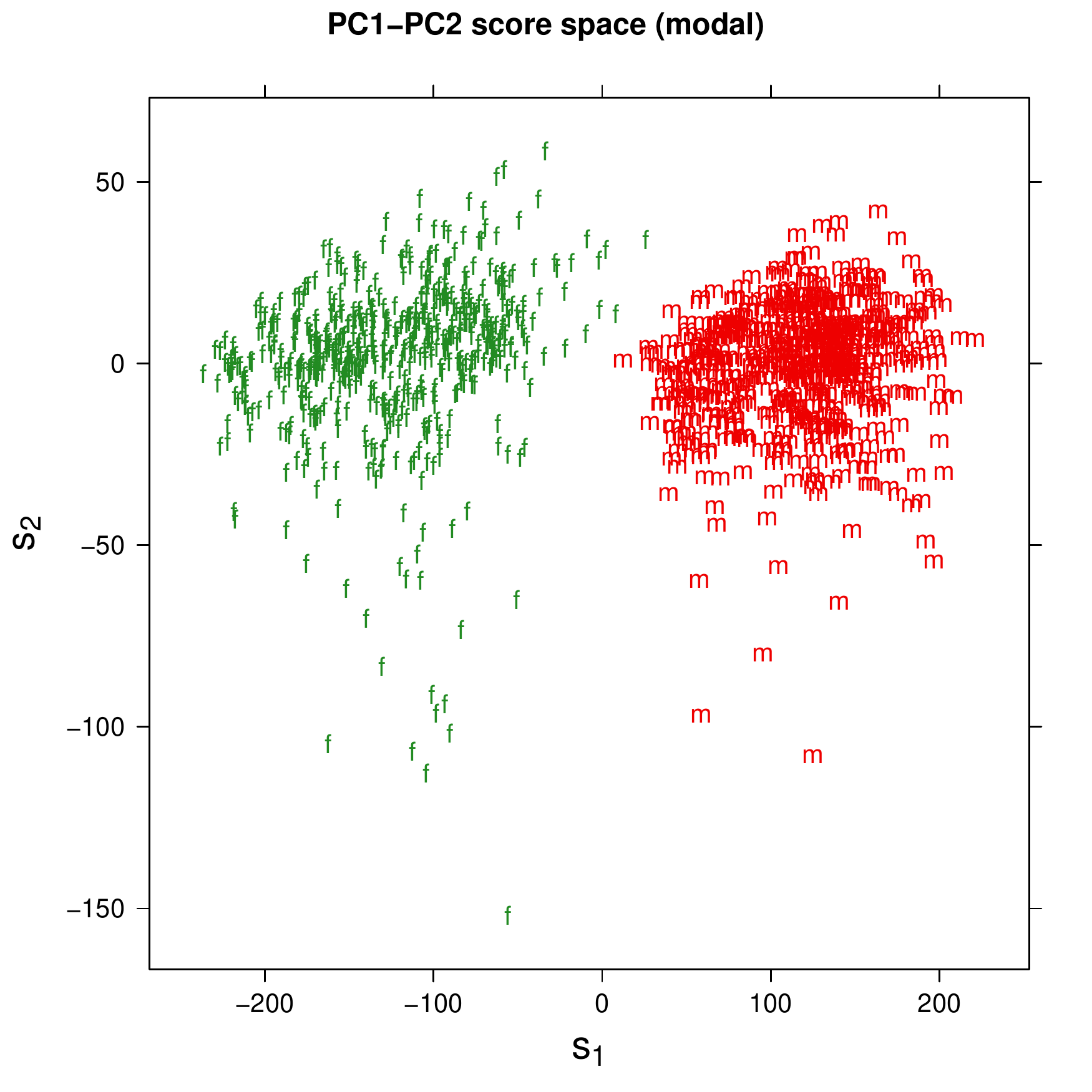}
\caption{Speakers’ modal PC1--PC2 score spaces. Letters \emph{f} and \emph{m} indicate PC scores (i.e. the $s_1$ and the $s_2$) of female and male voices, respectively.}
 \label{fig: fig5}
\end{figure}

Like Figure \ref{fig: fig2}, Figure \ref{fig: fig4} shows the effects of PC1 and PC2 functions on the mean f0 curve. However, in this case the curves present only modal speech from both female and male speakers. For this data set, PC1 already explains 81\% of pitch variation while PC2 explains only 2\%. Since the mean modal curves of female speakers evidently differ from those of male speakers (see Figure \ref{fig: fig1}), fPCA  captures a majority of this variation with the first PC score. In fact, the minus curve of the PC1 function in Figure \ref{fig: fig4} is highly similar to the modal mean f0 curve of male speakers and the plus curve similar to that of female speakers.
Figure \ref{fig: fig5} reveals the f0 differences between the sexes captured by the $s_1$ in the PC1--PC2 score space: female (\emph{f}) and male (\emph{m}) scores form their own distinct groups and are even more visibly separated than the scores of the three voice conditions. These observations suggest that swapping the $s_1$ across sexes will cause a major change in original f0 trajectories.

\section{Speaker de-identification methods} 
\label{SDM}

\subsection{Disguised voice}
\label{disguise}

\noindent As shown in Section \ref{fPCA}, fPCA revealed that positive and high PC1 scores are associated with intended child voice, which was found to be an effective voice disguise strategy against ASV systems \citep{gonzalez2019limits}. Based on this finding, modal voice was manipulated towards intended child voice by reconstructing the curves using fPCA components of both modal and intended child voices (see Equation \eqref{eq:f0-finite-expansion}). A similar reconstruction technique for f0 curves was originally proposed by the study of Gubian, Cangemi and Boves (2010).

To reconstruct manipulated versions of modal f0 curves, each speaker's $s_1$ from each modal curve were replaced with the mean $s_1$, which was calculated from the same speaker's intended child curves. Absolute values were used to calculate the mean to ensure that the $s_1$ will be a positive weighting (see Figure \ref{fig: fig3}). The rest of the original scores were kept the same. Hence, modal f0 curves would be reshaped to have a closer resemblance to f0 curves of intended child voices. Finally, the reconstructed f0 trajectories were inserted into original modal utterances using pitch-synchronous overlap-and-add (PSOLA) in Praat.

The number of PCs impacts the precision by which f0 curves can be represented. Gubian, Cangemi and Boves (2010) used only the first two PCs to reconstruct f0 curves. However, in the AVOID data the first two PCs explain only approximately 50\%--70\% of variation (see Section \ref{fPCA}). To avoid reconstructing over-smoothed f0 curves, we selected the number of PCs that explains $> 90\%$ of variation. This corresponds to retaining the first 30 PCs. This decision was based on visual inspections of the curves and listening to de-identified samples. 

\begin{figure} [h!]
\centering
\includegraphics[width=0.7\textwidth, angle=0]{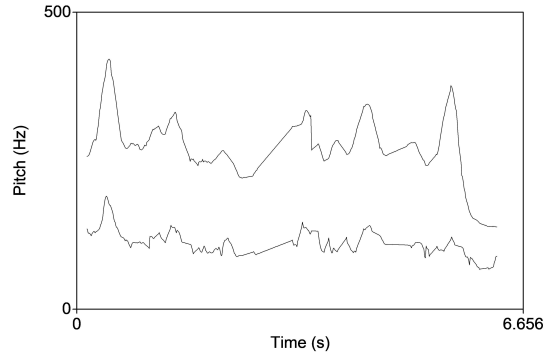}
\caption{Manipulated (above) and original (below) f0 trajectories extracted from the same speech utterance. The utterance \emph{Kun auringonvalo osuu sadepisaroihin ilmassa, ne käyttäytyvät kuin prismat ja muodostavat sateenkaaren} is spoken by a male speaker. The trajectories are interpolated for visual purposes.}
\label{fig: fig6}
\end{figure}

Figure \ref{fig: fig6} shows an example of an output of the pitch reconstruction. In this example, manipulated f0 trajectory (in black) is at substantially higher frequencies compared to the original (grey) trajectory. Intonational peaks are also more accentuated in the manipulated trajectory, which is a characteristic feature of intended child voice in the AVOID corpus. Nonetheless, some visible similarities remain in the shape of the manipulated trajectory since the intonational peaks of the two curves are time-aligned. This is due to fact that the rest of the original PC weightings were kept the same, even though the most influential PC weighting, the $s_1$, has been replaced.

It should be noted that the fPCA-based pitch reconstruction does not constantly yield substantially higher f0 trajectories compared to original trajectories as in Figure \ref{fig: fig6}, because the amount of increase depends on differences between speaker-dependent modal and disguised voices. Still, on average the AVOID speakers tend to use evidently higher-pitched voices when imitating children \citep{hautamaki2017acoustical}, which is reflected in manipulated trajectories.

\subsection{Cross-sex voice}\label{cross-sex}

\noindent Although transferring primary modes of pitch variation from one speaking style to another provides a novel approach to pitch manipulation, from a practical point of view disguised speech from the same speakers is rarely available for speaker de-identification. To demonstrate the usability of fPCA-based pitch manipulation without disguised speech, modal f0 curves of opposite sexes were also used as a model for speaker de-identification. In other words, f0 trajectories were manipulated using cross-sex f0 variation instead of disguised voices. Although this study is plausibly the first attempt to anonymize f0 trajectories using fPCs of cross-sex voices, simple linear transformations of cross-sex f0 has been previously implemented to improve speaker de-identification \citep{champion2021study}.

The f0 manipulation process using cross-sex PC scores was similar as with the intended child voices (see Section \ref{disguise}), although in this case the original $s_1$ was replaced with the mean $s_1$ from the curves of the opposite sex. The rest of the PC scores were unchanged. As a result, f0 manipulation using the mean of $s_1$ from cross-sex f0 curves raised male and lowered female speakers' pitch.

\subsection{Formant shifts}\label{formant}

\noindent Although manipulation of f0 affects human perception of a speaker's voice, for the state-of-the-art ASV systems formant manipulation is expected to be a more efficient speaker de-identification method. This is because ASV systems use spectral features to parametrize speech signals. Still, as far as the authors are aware, recent literature on speaker de-identification lacks clear empirical evidence regarding the effects of constant shifts of formant frequencies on the state-of-the-art ASV performance. Therefore, in addition to the fPCA-based pitch reconstruction, frequencies of the first three formants were raised by 10\% to 20\% using the Vocal toolkit \citep{PVT} function \emph{Change formant}, which modifies formant frequencies based on the Burg method provided in Praat. To be consistent with pitch manipulations based on child mimicking, formants were raised, rather than lowered\footnote{Yet, in \citep{hautamaki2017acoustical} formants (F1-F4) demonstrated somewhat inconsistent shifts between normal and intended child voices: approximately 30\% of utterance pairs showed no differences between the two voice conditions, while for approximately 70\% of utterance pairs, formants increased and decreased inconsistently.}. As a result, all modal utterances in speech data were anonymized by increasing the frequencies of the first three formants and by replacing the f0 trajectory as explained in Sections \ref{disguise} and \ref{cross-sex}.

The following four steps summarize the whole proposed speaker de-identification process:

\begin{enumerate}
  \item Extract f0 trajectories from original and modelling utterances, which are used for manipulation, and transform them to continuous curves.
  \item Perform fPCA on the curves and collect PC scores.
  \item For each original utterance, replace the first PC score with the mean first score from the modelling utterances, reconstruct the curves and change the original f0 trajectory to the modified one. Since the resulting f0 manipulation is solely based on the modelling utterances, it does not require extensive amounts of training data.
  \item Finally, increase the first three formant frequencies by 10\%-20\%.
\end{enumerate}
These steps can be implemented by adapting our publicly available code\footnote{Our code is publicly available at \url{https://github.com/laurivel/de_id-FDA-f0-formants}}.

\section{De-identification validation} \label{vali}

\subsection{Automatic speaker verification} \label{ASV}
\noindent To test the performance of the proposed speaker de-identification techniques, a widely used ASV approach using the Kaldi toolkit \citep{povey2011kaldi} was applied. Kaldi was also used to implement objective speaker verifiability metrics in the VoicePrivacy challenge \cite{tomashenko2020introducing}. The ASV approach is based on a deep neural network architecture for the extraction of x-vectors embeddings \citep{snyder2017deep,snyder2018xvector}. The x-vector system is built using a pre-trained augmented VoxCeleb 1 and 2 \citep{voxcelebCSL} data model recipe, which is available at \url{http://kaldi-asr.org/models/m7}. All speech recordings were downsampled to 16 kHz. For feature extraction configuration, 23-dimensional Mel-frequency cepstral coefficients (MFCCs) were extracted from 25 ms long frames every 10 ms. A short-time cepstral mean subtraction was applied over a 3-second sliding window and energy-based speech activity detection was used to drop the non-speech frames. For speaker similarity scoring, a probabilistic linear discriminant analysis (PLDA) \citep{PLDA_Original} was used where the extracted 512-dimensional x-vectors were centred, whitened, and unit length normalized.

The speaker verification setup for genuine and impostor speakers was built as follows: The speakers' modal voice segments of the first session were used as the enrollment material, resulting in phrase-dependent models per each speaker. The same speaker (genuine) trials correspond to the second session segments of the same speaker and, for different speaker (impostor) trials, the speech segments from the second session from the rest of the speakers were used. Thus, the speakers' original voice was considered as a point of reference, against which the anonymized test segments were compared. For each condition, the genuine and impostor trials were evaluated separately for female and male speakers. The number of genuine trials corresponding to female and male speakers were 620 and 580, respectively, while the corresponding numbers of impostor trials were 9300 and 8120.

Finally, the accuracy of the ASV system against anonymized speech was measured. While the enrollment data consists of natural (non-anonymized) speech, the de-identification methods are applied only to the test segments. Identifiability of speakers (whether natural or anonymized) is reported using \emph{equal error rate} (EER), a standard performance measure of ASV systems. The EER is an error rate obtained by equating false acceptance rate (FAR) and false rejection rate (FRR) by adjusting a detection threshold. The FAR is defined as the number of false acceptances divided by a number of impostor trials and the FRR is defined as the number of false rejections divided by the number of genuine trials. For any given de-identification method, equivalent processing steps are applied to both genuine and impostor trials. The higher the EER is for the anonymized trials, the more effective the de-identification performance is.

\subsection{Objective intelligibility} \label{intel}
\noindent Besides concealing speaker identities, another important aim in speaker de-identification is to maintain speech intelligibility. There are different approaches to intelligibility assessment with known trade-offs. Since the authors do not have access to large-scale crowdsourcing with Finnish listeners, we opted for objective measures. Instead of time-consuming listening tests, previous studies have proposed various metrics such as word error rate based on ASR systems along with different objective intelligibility measures \citep{spille2018predicting}. 
At the early stages of our study, we attempted to use a commercial ASR systems (e.g. Google Speech Cloud ASR) but the word recognition accuracy was quickly deemed insufficient for our data, possibly due to the nonstandard pronunciation in the disguised utterances. Thus, to avoid relying on the performance of ASR systems, in this study \emph{short-time objective intelligibility} (STOI) \citep{taal2010short,taal2011algorithm} was used to evaluate anonymized speech intelligibility. STOI is based on an intermediate measure for time-frequency regions with a discrete Fourier transform-based time-frequency-decomposition, and its relation with intelligibility has been confirmed using various human listening tests \citep{taal2010short,taal2011algorithm}. STOI compares a reference utterance (here, original modal speech utterance) with a processed utterance (here, anonymized version of the same utterance) and outputs a value between 0 and 1, which correlates with intelligibility. The edge case 1 correspond to perfect intelligibility, obtained on a pair of identical waveforms. Even if the interpretation of other values can be less straightforward, STOI is nonetheless useful as a relative measure to compare different waveform manipulation methods. The higher the value, the more intelligible speech is, on average terms. In other words, decreasing STOI can be interpreted as gradually increasing number of unrecognized words in speech.\footnote{A small example set of de-identified utterances and their STOI values are available at \url{https://github.com/laurivel/de_id-FDA-f0-formants}}

In this study, STOI was calculated using \texttt{Pystoi} \citep{STOI}. Although the STOI has been originally used to evaluate time-frequency weighted noisy speech, it has also been utilized in assessment of pathological speech \citep{janbakhshi2019pathological} and in speaker de-identification \citep{hashimoto2016privacy} studies. A restriction for measuring STOI is the requirement of time-aligned speech signals. This is, however, the baseline in speaker de-identification studies whenever the duration of original speech remains unmodified. For the proposed speaker de-identification methods, duration of speech was retained, and STOI measures were computed by comparing the original (unmodified) utterance with its anonymized version. Finally, an average STOI was calculated for each manipulated data set.

\section{Results} \label{Res}

\subsection{Automatic speaker verification}
\label{eer_r}

\noindent The effects of eleven different combinations of formant and f0 modifications on the performance of the x-vector-based ASV system are summarized in Table \ref{table:1}. The voice privacy protection against ASV was evaluated as percentage of EER (see Section \ref{ASV}), which is expected to increase when speaker de-identification methods are applied. As shown in Table \ref{table:1}, without de-identification (\emph{no anonymization}) the performance of the ASV system is almost perfect: EERs(\%) for male and female speakers are 0.32\% and 0\%, respectively. In addition, Table \ref{table:1} shows the effect of voice disguise, or speakers' intended child voices, on the ASV performance; EER(\%) increases from 0.32 to 11.94 for female speakers and from 0.00 to 15.17 for male speakers revealing the negative effect of voice disguise on ASV accuracy.

\begin{table}[hbt!]
\begin{center}
\caption{ASV performance in terms of EER(\%). Speaker de-identification (SDI) methods f0$_{D/S}$--F1-3$_{10/20}$ indicate fPCA-based pitch reconstruction and an increase of the first three formants. \emph{D} and \emph{S} present f0 manipulations based on child mimicry (class = disguise model) and opposite sex (cross-sex model). F1-3$_{10/20}$ equal constant increases of the first three formants by 10\% and 20\%, while f0$_{15}$ equals a constant increase of f0 by 15\% (class = reference). The highest EERs are marked using bold font.}
\label{table:1}
\begin{tabular}{ |c|c|c|c| }
 \hline\hline
 \textbf{class} & \textbf{SDI} & \textbf{EER (female)} & \textbf{EER (male)} \\
 \hline\hline
 \multirow{7}{7em} {reference} & no anonymization & 0.32 & 0.00\\
 & natural voice disguise & 11.94 & 15.17\\
 & F1-3$_{10}$ & 12.42 & 11.21\\
 & F1-3$_{20}$ & 13.55 & 15.86\\  
 & f0$_{15}$ & 0.32 & 0.17\\ 
 & f0$_{15}$--F1-3$_{10}$ & 10.00 & 10.34\\
 & f0$_{15}$--F1-3$_{20}$ & 12.42 & 15.69\\
 \hline
 \multirow{3}{7em}{disguise model} & f0$_{D}$ & 0.48 & 0.34\\
 & f0$_{D}$--F1-3$_{10}$ & 8.87 & 11.38\\  
 & f0$_{D}$--F1-3$_{20}$ & 13.87 & 16.03\\ 
 \hline
 \multirow{3}{7em}{cross-sex model} & f0$_{S}$ & 0.48 & 0.17\\
 & f0$_{S}$--F1-3$_{10}$ & 11.61 & 12.07\\ 
 & \textbf{f0$_{S}$--F1-3$_{20}$} & \textbf{16.94} & \textbf{17.24}\\  
 \hline
\end{tabular}
\end{center}
\end{table}

The results in Table \ref{table:1} confirm that fPCA-based pitch reconstruction (f0$_{D/S}$) negatively affects the ASV accuracy. However, the effect is weak because for female speakers the EER increases only from 0.32\% to 0.48\%. Similarly, for male speakers the increase is from 0.00\% only to 0.17\%--0.34\% depending on the model data (\emph{D} and \emph{S}). Simple formant shifts (F1-3$_{10/20}$) are more effective in deceiving ASV as EER(\%) increases to 12.42--13.55 for female and to 11.21--15.86 for male speakers. As expected, shifting formant frequencies 20\% higher will increase the EER more than shifting them 10\% as it manipulates the original spectral features further than the latter.

For speaker de-identification, merging 20\% formant shifts to f0 manipulation using the fPCA-based cross-sex model yielded the best result, which is indicated by bold font in Table \ref{table:1}.\footnote{In addition to increasing formant frequencies based on intended child speech, we performed a more anatomically motivated de-identification procedure for female speakers, where we used the original cross-sex FDA model for f0 but this time lowered the formant frequencies. However, the EERs were low compared to proposed method: the EERs were 1.45\% and 4.84\% when the formant frequencies were lowered by 10\% and 20\%, respectively. The STOI values were 0.87 and 0.83.} The same method, but using the fPCA-based disguise model instead of cross-sex model for f0 manipulation, resulted in the second highest EERs. Therefore, both the f0$_{S}$--F1-3$_{20}$ and the f0$_{D}$--F1-3$_{20}$ approaches resulted in higher EER(\%)s than natural voice disguise and outperformed the reference methods F1-3$_{20}$ and f0$_{15}$--F1-3$_{20}$. Surprisingly, for speakers of both sexes F1-3$_{20}$ yielded higher EER(\%)s than f0$_{15}$--F1-3$_{20}$, where in addition to the same formant shifts, the original f0 was increased by a constant 15\%. In addition, using f0$_{D}$--F1-3$_{10}$ for female speech yielded lower EER(\%) than F1-3$_{10}$ or f0$_{15}$--F1-3$_{10}$, even though using f0$_{D}$--F1-3$_{20}$ outperformed both F1-3$_{20}$ and f0$_{15}$--F1-3$_{20}$. Therefore, the connection between f0 modifications and the change of EERs can be seen to exhibit partial incoherence.

Despite the incoherence in the effects of f0 manipulations on EERs regarding female speakers, the results indicate that the more speech features are modified, the more challenging it is to identify speakers. Speech modifications, however, will eventually affect the intelligibility of speech, which is discussed in the next section.

\subsection{Objective intelligibility} \label{stoi_r}

\noindent Intelligibility of anonymized speech was evaluated using the STOI measure (see Section \ref{intel}). The results are presented in Table \ref{table:2}, in which the speaker de-identification techniques are presented similarly as in Table \ref{table:1}. Overall, all techniques yielded moderate intelligibility as the mean STOI values ranged from 0.63 to 0.94; if the compared pair of speech signals would be exactly the same signals, STOI would be 1.00. 

\begin{table}[hbt!]
\begin{center}
\caption{Mean STOI measures for anonymized speech. The numbers in parentheses show minimum and maximum values. Speaker de-identification (SDI) methods f0$_{D/S}$--F1-3$_{10/20}$ indicate fPCA-based pitch reconstruction and an increase of the first three formants. \emph{D} and \emph{S} present f0 manipulations based on child mimicry (class = disguise model) and opposite sex (cross-sex model). F1-3$_{10/20}$ equal constant increases of the first three formants by 10\% and 20\%, while f0$_{15}$ equals a constant increase of f0 by 15\% (class = reference). The SDI method that produced the highest EERs is marked using bold font.}
\label{table:2}
\begin{tabular}{ |c|c|c|c| }
 \hline\hline
 \textbf{class} & \textbf{SDI} & \textbf{STOI (female)} & \textbf{STOI (male)}  \\
 \hline\hline
 \multirow{5}{4em}{reference} & F1-3$_{10}$ & 0.82 (0.64-0.96) & 0.80 (0.58-0.96)\\
 & F1-3$_{20}$ & 0.76 (0.56-0.92) & 0.69 (0.51-0.85)\\ 
 & f0$_{15}$ & 0.86 (0.75-0.96) & 0.94 (0.87-0.97)\\ 
 & f0$_{15}$--F1-3$_{10}$ & 0.73 (0.59-0.87) & 0.76 (0.56-0.91)\\
 & f0$_{15}$--F1-3$_{20}$ & 0.68 (0.51-0.82) & 0.67 (0.49-0.81)\\
 \hline
 \multirow{3}{4em}{disguise model} & f0$_{D}$ & 0.80 (0.66-0.92) & 0.88 (0.73-0.95)\\
 & f0$_{D}$--F1-3$_{10}$ & 0.69 (0.55-0.85) & 0.71 (0.53-0.86)\\ 
 & f0$_{D}$--F1-3$_{20}$ & 0.63 (0.48-0.78) & 0.64 (0.48-0.78)\\
 \hline
 \multirow{3}{4em}{cross-sex model} & f0$_{S}$ & 0.90 (0.83-0.96) & 0.87 (0.75-0.93)\\
 & f0$_{S}$--F1-3$_{10}$ & 0.78 (0.63-0.88) & 0.71 (0.52-0.87)\\ 
 & \textbf{f0$_{S}$--F1-3$_{20}$} & \textbf{0.71 (0.53-0.82)} & \textbf{0.64 (0.47-0.82)}\\ 
 \hline
\end{tabular}
\end{center}
\end{table}

The highest STOI mean values for female (0.90) and male (0.94) speakers were achieved by simply using f0 manipulations (f0$_{S}$ and f0$_{15}$), which, however, yielded poor EER(\%)s considering speaker de-identification (see Section \ref{eer_r}). On the other hand, mere formant shifts, F1-3$_{10/20}$, which produced better voice privacy regarding the ASV performance, had a worse effect on the STOI: for female and male speakers, the F1-3$_{10}$ reduced STOI to 0.82 (female) and to 0.80 (male) while the F1-3$_{20}$ caused the STOI to decline to 0.76 (female) and to 0.69 (male). The f0$_{D}$--F1-3$_{20}$, which yielded the highest EER(\%), produced the lowest mean STOI for male speakers (0.64) and the fourth to last mean STOI (0.71) for female speakers.

Overall, the results shown in Tables \ref{table:1} and \ref{table:2} demonstrate that it plausible to manipulate f0 and formant frequencies to achieve decent speaker de-identification, but not without a cost in intelligibility. Depending on the aims of speech modifications, either the strength of the anonymization via major speech modifications or retaining high speech intelligibility via minor modifications might be more preferable.

\subsection{De-identification against more informed adversary}

\noindent Up to this point, our focus has been on de-identification methods, using an off-the-shelf ASV system to evaluate efficacy of de-identification. By viewing the de-identification and ASV methods, respectively, as proxies for `defense' (aimed at privacy protection) and `attack' (aimed at breaching the privacy) sides, the assumptions of the above ASV set-up can be summarized as follows.
    \begin{itemize}
        \item The attacker knows the target (hypothesized) speaker. In particular, the attacker has prior data of the speaker (the enrollment utterances);
        \item The attacker has access to an unlabeled new recording (the test utterance) and wishes to establish whether or not the speaker is the same as the hypothesized speaker; 
        \item The attacker does \textbf{not} know the details of the de-identification techniques. 
    \end{itemize}
The last point is implied by noting that the ASV training data (VoxCeleb) consists of natural data only and with disjoint set of speakers with the AVOID corpus.  

Nonetheless, one may argue that a more informed attacker could have prior knowledge of the language, the de-identification techniques, or both. In our last experiment, we consider such an informed adversary. In practice, we apply PLDA domain adaptation recipe available in the Kaldi toolkit (see also \cite{Bousquet2019}). The speakers used for domain adaptation and evaluation are disjoint. Since the above ASV experiments exhausted all the 60 speakers, we designed another protocol for this experiment. To this end, a total of 20 AVOID speakers (10 per gender) is used for domain adaptation while the remaining 40 AVOID speakers (21 female and 19 male) are held out for evaluation. The domain adaptation data (total 2600 utterances per gender\footnote{20 sentences (10 from each of two sessions) $\times$ 10 speakers $\times$ (no anonymization + natural disguise + 11 de-identification methods) = 2600.}) consists of natural utterances and samples of \emph{all} the de-identification methods considered above. This models a situation where the adversary has prior information of the de-identification methods in terms of speech examples (of speakers other than the hypothesized speaker).

The results shown in Table \ref{table:domain-adapt-female} indicate decreased EERs for the case of domain-adapted ASV models, in most cases. This suggests less effective privacy protection against informed attackers, as expected. Nonetheless, the relative efficacy of alternative de-identification methods remains the same as above expect for female speakers' natural disguise, which now yields the highest EER (9.76\%). For male speakers, fPCA-based pitch manipulations with the formant shifts still result in the best de-identification (6.58\%).

\begin{table}[hbt!]
\begin{center}
\caption{The impact of less- vs. more informed adversary to speaker de-identification performance. Here, \textbf{ASV1} is a pre-trained ASV model (no data from the AVOID corpus), while \textbf{ASV2} is domain-adapted with 20 AVOID speakers (using natural and all de-identification samples). The highest EERs are marked using bold font.}
\label{table:domain-adapt-female}
 \begin{tabular} {|c|c|c|c|c|c|}
 \hline\hline
 \textbf{class} & \textbf{SDI} & \multicolumn{2}{|c|} {\textbf{female EER(\%)}} & \multicolumn{2}{|c|} {\textbf{male EER(\%)}}\\
  &  & \textbf{ASV1} & \textbf{ASV2} & \textbf{ASV1} & \textbf{ASV2}\\
 \hline\hline
 \multirow{7}{7em} {reference} & no anonym. & 0.48 & 0.48 & 0.00 & 0.53\\
 & \textbf{natural disguise} & 13.33 & \textbf{9.76} & 11.32 & 6.31\\
 & F1-3$_{10}$ & 12.38 & 3.57 & 12.89 & 2.37\\
 & F1-3$_{20}$ & 13.33 & 4.52 & 17.37 & 4.74\\  
 & f0$_{15}$ & 0.24 & 0.24 & 0.26 & 0.26\\ 
 & f0$_{15}$--F1-3$_{10}$ & 10.48 & 3.33 & 12.11 & 2.37\\
 & f0$_{15}$--F1-3$_{20}$ & 12.42 & 5.00 & 17.11 & 5.79\\
 \hline
 \multirow{3}{7em}{disguise model} & f0$_{D}$ & 0.71 & 0.48 & 0.53 & 0.26\\
 & f0$_{D}$--F1-3$_{10}$ & 9.04 & 3.10 & 13.16 & 3.42\\  
 & \textbf{f0$_{D}$--F1-3$_{20}$} & 13.33 & 5.95 & 17.11 & \textbf{6.58}\\ 
 \hline
 \multirow{3}{7em}{cross-sex model} & f0$_{S}$ & 0.48 & 0.24 & 0.26 & 0.53\\
 & f0$_{S}$--F1-3$_{10}$ & 11.67 & 3.10 & 13.95 & 3.42\\ 
 & \textbf{f0$_{S}$--F1-3$_{20}$} & \textbf{15.00} & 5.95 & \textbf{18.42} & \textbf{6.58}\\  
 \hline
\end{tabular}
\end{center}
\end{table}


\section{Discussion}
\label{Diss}
\noindent Although x-vector-based resynthesis is more secure speaker de-identification technique than modification of f0 and formant features, it can be highly complex and require (pre)trained models and speech technology tools, which are often lacking from languages other than the most spoken ones. Already using ASR systems in anonymization is problematic because the performance of ASR systems for Finnish, and for many other languages, is still far away from perfect, especially for speech that is not spoken in standard language. This paper has shown that FDA and particularly fPCA provides a novel and valid addition to formant-based speaker de-identification. Although the weightings of the first PC function of disguised and cross-sex speech were used as a basis of pitch manipulations, f0 trajectories can be manipulated based on any speech material including different languages. However, considering the fact that intonation has essential linguistic functions and it is highly dependent on the language \citep{de1998intonation}, using different languages to manipulate, for example, Finnish speech can affect speech quality. Yet, altering the weightings of the first PC function while preserving the rest of the original weightings will ensure that reconstructed f0 trajectories will diverge from but also fit properly to original speech resulting in anonymized and intelligible speech. Because the fPCA-based technique manipulates f0 using dynamic features from other utterances, original f0 trajectories cannot be restored by applying constant pitch shifts.

The proposed data-driven speaker de-identification method does not require extensive amounts of training data. The only limit is that the modelling speech should differ enough in f0 in comparison to the original speech: the more f0 trajectories differ, the more f0 trajectories are modified. Furthermore, depending on resource availability for x-vector-based resynthesis, fPCA-based pitch manipulation could be used with the x-vector approach instead of formant modifications to anonymize speech.

Performing FDA is also phonetically informative because it reveals main characteristics of pitch variation from data sets. As shown in Figures \ref{fig: fig1} and \ref{fig: fig2}, using intended child speech as a voice disguise strategy yields a high pitch in comparison to modal speech, and modal female speech exhibits a higher pitch than modal male speech, as expected. It should be noted that FDA in this study focused on mean curves of voice conditions while examining curves of individual speakers might have revealed more details about the relation of speaker-dependent and disguise-dependent f0 variation. Using FDA for modelling voice disguise strategies and speaker-dependent f0 curves would be a potential topic for future work.

\section{Conclusion}
\noindent This study has demonstrated how applying FDA on f0 trajectories can improve speaker de-identification in a data-driven and phonetically controllable manner. In addition to a simple increase of the first three formant frequencies, f0 trajectories were reconstructed using functional principal components, which were calculated from two different kinds of data sets: 1) disguised and 2) cross-sex speech. The aim was to explore a novel data-driven approach to speaker de-identification, which requires no large speech corpora nor pre-trained language models. The effectiveness of the speaker de-identification processes were evaluated with two objective measures, EER(\%) and the STOI. EER(\%) showed reasonable voice privacy protection against an x-vector-based ASV system as the proposed methods deteriorated the ASV performance from 0\% to 17\% in terms of EERs. Although the degradation was primarily due to formant shifts, adding FDA-based f0 manipulation to formant shifts increased relative EERs 9\% to 25\%. In addition, STOI measures indicated relatively high intelligibility for anonymized speech using the most effective techniques (STOI = 0.64--0.71). As expected, comparing EERs and STOI values indicated that intensive speech manipulations will eventually compromise speech intelligibility.

\section*{Acknowledgements}

\noindent This work was supported by the Academy of Finland (Proj. No. 309629 --- entitled “NOTCH: NOn-cooperaTive speaker CHaracterization”).

\bibliography{mybibfile}

\end{document}